\documentclass[10pt]{article}
\usepackage{epsf}
\newcommand{\dd}{\mbox{d}}

\def\overlay#1#2{\ifmmode%
\setbox0=\hbox{$#1$}%
\setbox1=\hbox to\wd0{\hss$#2$\hss}\else%
\setbox0=\hbox{#1}%
\setbox1=\hbox to\wd0{\hss#2\hss}\fi%
#1\hskip-\wd0\box1 }
\begin{document}
\begin{titlepage}
\title{Quantum Field Theory \\ of \\ Vortices in Superfluid Films}
\author{J\"org P.\ Kottmann and Adriaan M.\ J.\ Schakel \\ Institut f\"ur
Theoretische Physik \\ Freie Universit\"at Berlin \\ Arnimallee 14, 14195
Berlin \\ e-mail: schakel@physik.fu-berlin.de }
\date{November 29, 1997}
\maketitle
\begin{abstract}
A quantum field theory, consisting of the effective action of sound waves
linearly coupled to a Chern-Simons term, is proposed to describe the
dynamics of vortices in a superfluid film at the absolute zero of
temperature.  
\end{abstract}
\end{titlepage}
In a superfluid film at the absolute zero of temperature, vortices are
induced by quantum fluctuations.  These vortices are to be distinguished
from the thermally induced ones at finite temperature.  Whereas the latter
are classical objects, the zero-temperature vortices are quantum objects.
Besides in superfluids, quantum vortices play an important role in various
other two-dimensional quantum systems such as the fractional quantized Hall
effect, superconducting films and Josephson junction arrays at the absolute
zero of temperature.

Since zero-temperature vortices in a superfluid film are point-like quantum
objects, one may wonder whether their dynamics can be described by a quantum
field theory.  We shall argue in this Letter that this is indeed the case.
Because at zero temperature the normal fluid can be ignored, an important
result of Helmholtz applies, stating that a vortex moves with the fluid,
i.e., at the local velocity of the fluid.  This theorem, implying that the
dynamics of vortices is determined by the superfluid component, provides a
stringent constraint on the quantum field theory we are seeking.  An other
constraint follows from the observation due to Haldane and Wu \cite{HW} that
when a vortex encircles a boson, it accumulates a geometric phase of $2
\pi$.  It will be shown that in two dimensions these requirements can be
fulfilled by a Chern-Simons theory, one of whose main characteristics is
that it involves a vector field which has no independent dynamics.

Our approach is to be contrasted with other modern treatments of the
dynamics of vortices in superfluid films \cite{Popov,Hatsuda,Lee,Arovas}.
These theories give a quantum-mechanical description of vortices with delta
functions representing their worldlines.  In our quantum field approach, on
the other hand, vortices are described by a nonsingular quantum field.

The theory we propose to describe sound waves as well as vortices in a
superfluid film at zero temperature is given by the Lagrangian
\begin{equation} \label{vbn} 
{\cal L} = {\cal L}_{\rm eff} - e n a_{0} + \frac{e}{c} {\bf j} \cdot
{\bf a} + {\cal L}_{{\rm CS}}.
\end{equation} 
It consists of the effective Lagrangian ${\cal L}_{\rm eff}$ describing
the superfluid without vortices, {\it linearly} coupled via the particle
number current $(n,{\bf j})$ to a vector field $(a_0,{\bf a})$ governed
by a Chern-Simons term
\begin{equation}
{\cal L}_{{\rm CS}} = \frac{1}{2 c} {\bf a} \times \partial_{0} {\bf a} -
a_{0} {\bf \nabla} \times {\bf a},
\end{equation} 
with $c$ the sound velocity.  The vector field, which has no independent
dynamics, accounts for the vortices, with the Chern-Simons term encoding
the geometric phase acquired by a vortex when it winds around a boson.
The quantum-mechanical analog of such a term, representing the linking
number of a closed boson and vortex trajectory \cite{Kleinert}, was
introduced in the problem by Arovas and Freire \cite{Arovas}.  The
charge $e$ appearing in (\ref{vbn}) will be determined shortly.  The
effective Lagrangian ${\cal L}_{\rm eff}$ is given by
\begin{equation} \label{ok}
{\cal L}_{{\rm eff}} = -\bar{n}\left[\hbar\partial_{0}\varphi +
\frac{1}{2m}(\hbar {\bf \nabla} \varphi)^{2} \right] + \frac{\bar{n}}{2m
c^{2}}\left[\hbar\partial_{0}\varphi + \frac{1}{2m}(\hbar {\bf
\nabla}\varphi)^{2}\right]^{2},
\end{equation} 
where the dimensionless field $\varphi$ is the gapless Goldstone mode of
the spontaneously broken global U(1) symmetry, $m$ is the mass of the
Bose condensed atoms, and $\bar{n}$ is the particle number density of the
fluid at rest.  Physically, this nonlinear theory governs the sound
mode, with the dispersion $\omega^2 = c^2 {\bf k}^2$, where $\omega$ is
the (angular) frequency and ${\bf k}$ the wave vector.  The effective
theory can either be inferred from general symmetry arguments
\cite{GWW}, or explicitly derived from the microscopic Bogoliubov theory
by integrating out quantum fluctuations \cite{effBEC}.  It gives a
complete description of the superfluid valid at low energies and small
momenta.  The same effective theory appears in the context of (neutral)
superconductors and of classical hydrodynamics \cite{eff}.

The particle number current that follows from (\ref{ok}) reads
\begin{eqnarray}  \label{roh}
n &=& \bar{n} -\frac{\bar{n}}{m c^{2}} \left[\hbar \partial_{0} \varphi +
\frac{1}{2 m} (\hbar {\bf \nabla} \varphi)^{2}\right] \nonumber \\ 
{\bf j} &=& n {\bf v},
\end{eqnarray} 
where ${\bf v} = (\hbar/m) \nabla \varphi$ is the superfluid velocity field.
Physically, the first equation in (\ref{roh}) reflects Bernoulli's principle
which states that in regions of rapid flow, the density and therefore the
pressure is low.

To see under which circumstances the higher order terms in the effective
theory ${\cal L}_{\rm eff}$ can be neglected, we note that additional
derivatives are always accompanied by additional factors of $\hbar k/mc$,
with $k$ the wave number $k = |{\bf k}|$.  This means that third and higher
order in the field $\varphi$ can be ignored provided the wave number is smaller
than the inverse coherence length $\xi = \hbar/mc$
\begin{equation}  \label{zwd}
k < 1/\xi.
\end{equation} 
For $^4$He the coherence length, or Compton wavelength, is about 10 nm which
is of the order of the interatomic spacing.  In the experimentally
determined spectrum of superfluid $^4$He, the inverse of this value, 0.1
nm$^{-1}$, marks the point beyond which the spectrum ceases to be linear.
That is, the region defined by (\ref{zwd}) coincides with the region where
the superfluid $^4$He spectrum is linear and the description in terms of a
sound mode is applicable.

Before proceeding, we emphasize that the coupling of the Chern-Simons
vector field to the sound mode is linear and not minimal---as is usually
the case in gauge theories.  As a result, the coupled theory (\ref{vbn})
does not possess a gauge invariance involving a simultaneous local gauge
transformation of the Chern-Simons vector field and the matter field.
Due to the Higgs mechanism, a minimal coupling would inevitably result
in the disappearance of the gapless sound mode.  This would not be in
agreement with the physics we wish to describe in this Letter, which
is the dynamics of vortices in a {\it compressible} superfluid film.

We next consider the field equation for $a_0$.  As is common in theories
containing a Chern-Simons term, this component of the vector field merely
plays the role of a Lagrange multiplier.  We find
\begin{equation} \label{para}
\nabla \times {\bf a} = - e n,
\end{equation} 
or when integrated
\begin{equation} 
\Phi  = -e N,
\end{equation} 
where $N = \int \dd^2 x n$ is the particle number and $\Phi = \int \dd^2
x \nabla \times {\bf a}$ the ``magnetic'' flux associated with the
Chern-Simons vector field.  That is, particles carry besides a particle
number charge also a flux.  In the Coulomb gauge $\nabla \cdot {\bf a} =
0$, Eq.\ (\ref{para}) yields the solution
\begin{equation}  \label{aa}
{\bf a}(t,{\bf x}) = -e \nabla_{{\bf x}} \times \int
\dd^{2}x' G({\bf x} - {\bf x}') n(t,{\bf x'}) ,
\end{equation} 
where $G({\bf x})$ is the two-dimensional Green function of the Laplace
operator
\begin{equation} 
G({\bf x}) = - \frac{1}{2 \pi} \ln(|{\bf x}|),
\end{equation} 
i.e., $\nabla^2 G({\bf x}) = - \delta({\bf x})$, with $\delta({\bf x})$ the
two-dimensional delta function.  The solution shows that the Chern-Simons
vector field is entirely determined by the particle number density $n$.

To identify the flux carried by the particles, we calculate the Berry phase
$\gamma(\Gamma)$ by evaluating the Wilson loop $W(\Gamma) =\exp[i \gamma
(\Gamma)]$.  The latter is obtained by integrating the Chern-Simons
vector field ${\bf a}$ around a closed loop $\Gamma$ in the superfluid,
\begin{equation} \label{Berry}
\gamma(\Gamma) = \frac{e}{\hbar c} \oint_{\Gamma} \dd {\bf l}
\cdot {\bf a}(t,{\bf x}) = \frac{e^2}{h c} \int \dd^{2}x' \oint_{\Gamma}
\dd {\bf l} \cdot [ \nabla_{{\bf x}} \times \ln(|{\bf x} - {\bf x}'|)
n(t,{\bf x'}) ],
\end{equation} 
where we substituted the explicit form (\ref{aa}) for ${\bf a}$.  Provided
we choose $e^2 = h c$, this geometric phase coincides with the one obtained
by Haldane and Wu \cite{HW} who transported a vortex adiabatically around
the closed path $\Gamma$.  Their starting point was an {\it Ansatz} for the
multivortex wavefunction of the interacting Bose condensate.  They concluded
that, apart from corrections due to residual vortex interactions that become
small in the dilute-vortex limit, the right-hand side of (\ref{Berry})
yields $\pm 2\pi$ times the mean number of superfluid particles enclosed by
the closed loop $\Gamma$.  In other words, the vortex which was taken around
the closed path $\Gamma$ sees the encircled bosons as sources of geometric
phase.  In the theory proposed here, this counting is provided by the flux
imparted to a particle by the Chern-Simons term.

We next introduce external point vortices into the theory to determine
their action.  To this end, we consider only the following terms of the
Lagrangian (\ref{vbn}):
\begin{equation}  \label{rel}
{\cal L}_{\rm ext} = -\bar{n}\left[\hbar\partial_{0}\varphi + \frac{1}{2m}(\hbar
{\bf \nabla} \varphi)^{2} \right] + \frac{e}{c} {\bf j}\cdot {\bf a}_{\rm ext},
\end{equation} 
where the particle number current ${\bf j}$ reads ${\bf j} = \bar{n} {\bf v}$ in
this approximation, and where the Chern-Simons vector field ${\bf a}_{\rm
ext}$ is given by (\ref{aa}) with $n$ replaced by the external vortex
density
\begin{equation} 
n_{\rm ext}(t,{\bf x}) = \sum_\alpha w_\alpha \delta[{\bf x} - {\bf
X}^\alpha(t)].
\end{equation}
Explicitly, 
\begin{equation}  \label{aex}
{\bf a}_{\rm ext} (t, {\bf x}) = \frac{e}{2 \pi} \sum_\alpha w_\alpha
\nabla \arctan\left( \frac{x_2 - X_2^\alpha(t)}{x_1 - X_1^\alpha(t)}
\right),
\end{equation} 
where ${\bf X}^\alpha(t)$ denotes the location of the $\alpha$th vortex
with winding number $w_\alpha$.  The omitted terms in (\ref{rel}) are
all of higher order in derivatives which can be ignored at low energies
and small momenta.  The field equation for $\varphi$ obtained from
(\ref{rel}) can be easily solved to yield
\begin{equation} \label{feq}
\varphi(t,{\bf x}) = - \frac{e}{\hbar c} \int \dd^{2}x' G({\bf x} - {\bf
x}') \, \nabla_{{\bf x}'} \cdot {\bf a}_{\rm ext}(t,{\bf x'}) .
\end{equation}
When substituting this back into the Lagrangian (\ref{rel}), which is
tantamount to integrating out the phonons, we find for the action $S_{\rm
ext} = \int \dd^3 x {\cal L}_{\rm ext}$ describing the external charges:
\begin{equation} \label{action}
S_{\rm ext} = \bar{n} m \int \dd t\left[\frac{1}{2} \sum_\alpha \gamma_\alpha
{\bf X}^\alpha \times \dot{\bf X}^\alpha + \frac{1}{2\pi} \sum_{\alpha <
\beta} \gamma_\alpha \gamma_\beta \ln(|{\bf X}^\alpha - {\bf X}^\beta|)
\right],
\end{equation} 
where $\gamma_\alpha = (h/m) w_\alpha$ is the circulation of the
$\alpha$th vortex.  This action yields the well-known equations of
motion for point vortices in an incompressible two-dimensional fluid
\cite{Lamb,Lund}, which we thus have shown to be correctly reproduced by
the field theory proposed here.  The restriction to an incompressible
fluid, where the particle density $n$ is a constant $\bar{n}$, stems from
ignoring the higher-order derivative terms in (\ref{rel}).

Let us continue by calculating the scattering amplitude of two phonons
interacting via the Chern-Simons vector field at the tree level.  In the
frame where the sum of the two incoming momenta is zero, this corresponds to
the scattering of a phonon from a vortex.  An analogous situation arises in
the case of Aharonov-Bohm scattering, i.e., scattering of a massive
nonrelativistic particle from an infinitely thin magnetic flux tube.  It was
pointed out by Bergman and Lozano \cite{BL} that such a scattering process
can be described by a vacuum field theory consisting of a nonrelativistic
$|\psi|^4$-theory also coupled---albeit minimally---to a Chern-Simons term.

The scattering of sound waves from a vortex, formulated as a quantum
mechanical problem, was first studied by Pitaevskii in the Born
approximation \cite{Pitaevskii}.  (For a recent account and an extensive
list of references see Ref.\ \cite{Sonin1}, where also the close
connection with Aharonov-Bohm scattering is discussed.)  It was found
that in this approximation, the two-dimensional scattering amplitude
$f(\theta,k)$ is given by
\begin{equation}  \label{pit}
f(\theta,k) = \frac{1}{2} \sqrt{\frac{k}{2 \pi}} \frac{h}{ m c} {\rm e}^{i
\frac{\pi}{4}} \frac{\sin{\theta} \cos{\theta}}{1-\cos{\theta}} ,
\end{equation}  
where $\theta$ is the angle between the incoming and the scattered sound
wave and $k$ the wave number of both waves.  For small scattering angles,
the Born approximation breaks down.

At small energies and momenta, we can ignore the higher-order terms in the
Lagrangian (\ref{vbn}) and restrict ourselves to terms at most quadratic in
the phonon field:
\begin{eqnarray}   \label{theorie}
{\cal L}^{(2)} = \frac{\bar{n} }{mc^2} \Biggl\{&& \!\!\!\!\!\!\!\!\!\!\!\!
\frac{1}{2}(\hbar \partial_{0}\varphi)^2 - \frac{c^{2}}{2} (\hbar \nabla
\varphi)^2 + e a_{0} \left[\hbar \partial_0 \varphi + \frac{1}{2m} (\hbar \nabla
\varphi)^{2} \right] \nonumber \\ && \!\!\!\!\!\!\!\!\!\! - \frac{e\hbar^2}{mc}
\partial_{0} \varphi \nabla \varphi \cdot {\bf a}\Biggr\} + {\cal L}_{\rm CS} .
\end{eqnarray}
If we again introduce external vortices by replacing the Chern-Simons
vector field ${\bf a}$ with (\ref{aex}), the field equation for $\varphi$ becomes
\begin{equation} 
\partial_0^2 \varphi - c^2 \nabla^2 \varphi = 2 \frac{e}{mc} \partial_0 \nabla
\cdot {\bf a}_{\rm ext}
\end{equation} 
which is the equation found by Pitaevskii \cite{Pitaevskii} with the
contribution from the vortex motion ignored.

\begin{figure}
\begin{center}
\epsfxsize=12.cm
\mbox{\epsfbox{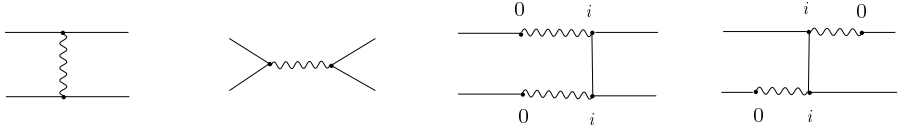}}
\end{center}
\caption{Tree graphs for the scattering of two phonons.}
\label{fig:tree}
\end{figure}

Let us return to the theory (\ref{theorie}) without external vortices.
In the gauge $\nabla \cdot {\bf a} = 0$, the nonzero components of the
Chern-Simons vector-field propagator are given by
\begin{equation}  
\raisebox{-0.2cm}{\epsfxsize=3.cm \epsfbox{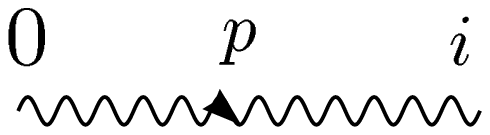} } \;\;\; : \;\;\; i
\hbar G_{i0}(p_{0},{\bf p}) = - i \hbar G_{0i}(p_{0},{\bf p}) = - \hbar
\epsilon_{ij} \frac{p_{j}}{{\bf p}^{2}},
\end{equation}  
while the phonon propagator reads
\begin{equation}  
\raisebox{-0.2cm}{\epsfxsize=2.7cm \epsfbox{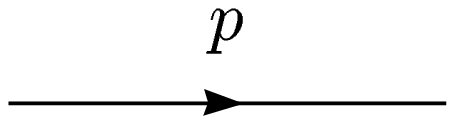} } \;\;\; : \;\;\; i
\hbar G(p_{0},{\bf p}) = \frac{m c^{2}}{\bar{n}} \frac{i
\hbar}{p_{0}^{2}-c^{2}{\bf p}^2 +i \eta} ,
\end{equation} 
where $\eta$ is a small positive constant that has to be taken to zero after
the loop integration over the energy $p_0$ has been carried out.  The
vertices of the theory are
\begin{eqnarray}   \label{een} 
\raisebox{-0.2cm}{\epsfxsize=3.cm \epsfbox{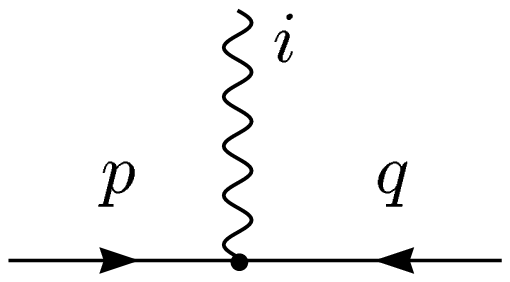} } &:& - \frac{i}{\hbar}
\frac{e\bar{n}}{m^2 c^{3}}(p_{0} q_{i} + p_{i} q_{0} ) \\ \label{zwee}
\raisebox{-0.2cm}{\epsfxsize=3.cm \epsfbox{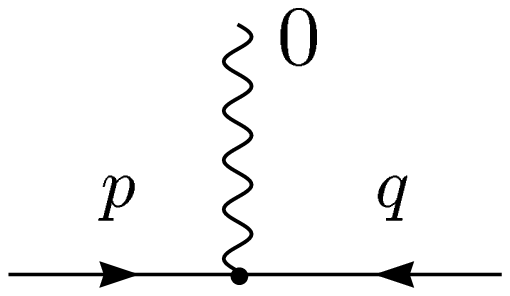} } &:& - \frac{i}{\hbar}
\frac{e \bar{n}}{m^2 c^{2}} {\bf p} \cdot {\bf q} \\ && \nonumber \\
\label{dree}  
\raisebox{-0.2cm}{\epsfxsize=3.1cm \epsfbox{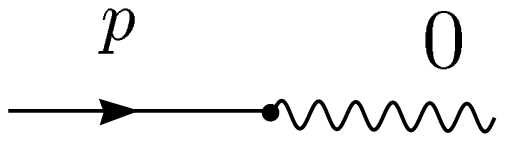} } &:& \frac{1}{\hbar}
\frac{e\bar{n}}{m c^{2}} p_{0}.
\end{eqnarray} 
The tree graphs for the phonon-phonon scattering are depicted in Fig.\
\ref{fig:tree}.
Of these, only the first graph survives; the second graph is identical zero,
while the third and fourth graph cancel each other.  Before evaluating the
first graph, let us pause for a moment and compare it with a similar graph
where instead of a Chern-Simons vector-field quantum, a phonon is exchanged.
The required vertices are contained in the higher-order terms of the theory
which have been ignored in (\ref{theorie}).  A dimensional analysis of the
type commonly used in the context of effective field theories
\cite{Weinberg}, shows that the exchange of a phonon can be ignored in
comparison to the exchange of a vector-field quantum provided
\begin{equation} 
k < \xi \bar{n},
\end{equation} 
where $k$ is the wave number of the incoming and outgoing phonons.  For
superfluid $^4$He this implies that the wave number should be smaller than
about 10 nm, which happens to coincide with the previous bound (\ref{zwd}).
This conclusion also applies to graphs involving an exchange of multiple
quanta.

Let us now evaluate the first graph in Fig.\ \ref{fig:tree}.  To
determine the normalization of the external lines we consider the energy
density ${\cal E}$ of the free phonon theory,
\begin{equation}  \label{Ham}
{\cal E}=\frac{\bar{n}}{2m c^{2}} \left[(\hbar \partial_{0} \varphi)^{2} + c^{2}
(\hbar \nabla \varphi)^2 \right].
\end{equation} 
The incoming and outgoing plane waves have the form
\begin{equation}  
\varphi(t,{\bf x}) = \zeta \left[e^{i(\omega t-{\bf k}\cdot {\bf x})}+
e^{-i(\omega t-{\bf k}\cdot {\bf x})}\right] ,
\end{equation} 
where $\zeta$ is the normalization constant which we fix by demanding that
the integrated energy density yields $\hbar \omega = \hbar c k$, the energy
of such a wave,
\begin{equation} 
\hbar \omega = \int \dd^{2} x {\cal E} = 2 \zeta^{2} \frac{\bar{n} \hbar^2
\omega^2}{m c^{2}} V,
\end{equation} 
where $V$ is the volume of the system.  In this way, we obtain for the
normalization constant
\begin{equation}  
\zeta=\sqrt{\frac{m c^2}{2 \hbar \omega \bar{n} V}} .
\end{equation} 
The first graph in Fig.\ \ref{fig:tree} is now easily evaluated, with the
result for the scattering amplitude $A(\theta,k)$
\begin{equation}    \label{treeres} 
\raisebox{-1.5cm}{\epsfxsize=3.cm \epsfbox{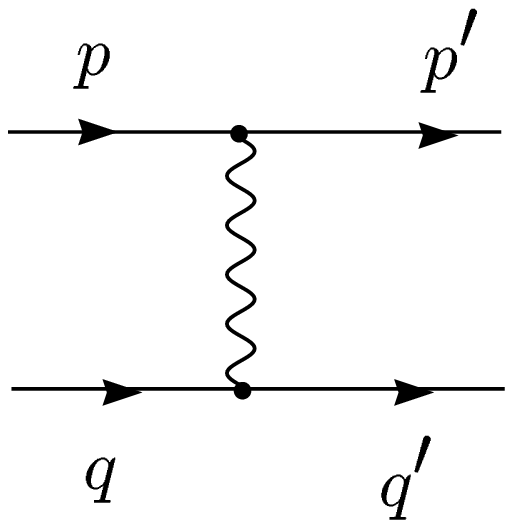} } \;\;\; : \;\;\; 
i A(\theta,k) =  -\frac{1}{2} \frac{h k}{m^2 c} \frac{\sin{\theta}
\cos{\theta}}{1-\cos{\theta}},
\end{equation} 
where in the frame chosen by us, the energy-momenta of the incoming and
outgoing phonons satisfy the equations
\begin{equation} 
p_0 = q_0 = p'_0 = q'_0, \;\;\; |{\bf p}|=|{\bf q}|=|{\bf p}'|=|{\bf q}'|,
\;\;\; {\bf p} = - {\bf q}, \; {\bf p}' = - {\bf q}'.
\end{equation} 
The scattering amplitude (\ref{treeres}) calculated here agrees with the
result (\ref{pit}) of Pitaevskii \cite{Pitaevskii}.  The difference in
kinematical factors in (\ref{pit}) and (\ref{treeres}) is due to a
different definition of the scattering amplitude in quantum mechanics
and in quantum field theory.

For identical particles, we have to add to the tree graph we just
calculated the graph with the two outgoing lines exchanged.  This yields
the same result as in (\ref{treeres}) with $\theta \rightarrow \theta -
\pi$.  Adding the two contributions, we find
\begin{equation} 
A^{\rm tot}(\theta,k) = i \frac{h k}{m^2 c} \cot \theta,
\end{equation} 
which diverges for both $\theta = 0$ and $\pi$.

A dimensional analysis of the loop corrections to this tree result reveals
that the expansion parameter is given by $(\xi k)^2$.  This is in agreement
with a conclusion of Sonin \cite{Sonin1}, which was based on the analogy
between the scattering of sound waves from a vortex and Aharonov-Bohm
scattering.  It shows that in the region where the field theory (\ref{vbn})
is applicable, the higher loop corrections are small.

In conclusion, we have shown that the quantum field theory proposed here to
describe zero-temperature vortices in a superfluid film, correctly
reproduces previously known results. In a future publication, we plan to
report the one-loop correction to the scattering amplitude considered here
and discuss the renormalizability of the theory.
\vspace{.5cm} \\
\noindent
{\bf Acknowledgments} \\
\noindent
The authors gratefully acknowledge discussions with D. Arovas and
H. Kleinert.  This work was performed as part of a scientific network
supported by the European Science Foundation, an association of 62 European
national funding agencies (see network's URL,
http://defect.unige.ch/defect.html).
\end{document}